\documentclass[conference]{IEEEtran}
\IEEEoverridecommandlockouts

\usepackage{cite}
\usepackage{amsmath,amssymb,amsfonts}
\usepackage{algorithmic}
\usepackage{graphicx}
\usepackage{textcomp}
\usepackage{verbatim} 
\usepackage{xcolor}
\def\BibTeX{{\rm B\kern-.05em{\sc i\kern-.025em b}\kern-.08em
    T\kern-.1667em\lower.7ex\hbox{E}\kern-.125emX}}

\usepackage{caption, subcaption}
\usepackage{url}

\usepackage{booktabs}
\usepackage{multirow}
\usepackage{siunitx}

\usepackage{enumitem}
\usepackage[english, status=draft, margin=false, inline=true]{fixme}
\fxusetheme{color}
\FXRegisterAuthor{jf}{ajf}{JF}

\makeindex

\begin{document}

\title{\emph{Fog Function}: Serverless Fog Computing for Data Intensive IoT Services}

\thispagestyle{plain}
\pagestyle{plain}


\author{\IEEEauthorblockN{Bin Cheng, Jonathan Fuerst, Gurkan Solmaz}
\IEEEauthorblockA{NEC Laboratories Europe, Heidelberg, Germany}
\and
\IEEEauthorblockN{Takuya Sanada}
\IEEEauthorblockA{NEC Solution Innovators, Ltd., Tokyo, Japan}
}


\maketitle

\begin{abstract}

Fog computing can support IoT services with fast response time and low
bandwidth usage by moving computation from the cloud to edge devices. However,
existing fog computing frameworks have limited \emph{flexibility} to support
dynamic service composition with a data-oriented approach.
Function-as-a-Service (FaaS) is a promising programming model for fog computing
to enhance flexibility, but the current event- or topic-based design of
function triggering and the separation of data management and function
execution result in \emph{inefficiency} for data-intensive IoT services. To
achieve both flexibility and efficiency, we propose a \emph{data-centric
programming model} called \emph{Fog Function} and also introduce its underlying
orchestration mechanism that leverages three types of contexts: data context,
system context, and usage context. Moreover, we showcase a concrete use case
for smart parking where Fog Function allows service developers to easily model
their service logic with reduced learning efforts compared to a static service topology. 
Our performance evaluation results show that the Fog Function can be scaled to
hundreds of fog nodes. Fog Function can improve system efficiency by saving
95\% of the internal data traffic over cloud function and it can reduce
service latency by 30\% over edge function. 

\end{abstract}


\section{Introduction}
\label{sec:introduction}

With the proliferation of IoT devices such as sensors, cars, drones, and
robots, these {\em IoT devices} not only produce lots of data but increasingly
consume the output of machine learning powered data processing pipelines to
take actions in a timely fashion~\cite{stoica2017berkeley}. Usually, data
producers and consumers are linked via {\em IoT services} that implement the
data processing logic to transform raw data into actionable results. Previously, central
clouds have been used as the main underlying infrastructure for
hosting such IoT services. However, due to the requirements of short response
time and low bandwidth use of many services associated with connected
vehicles, drones and cameras, there is a strong need to move data processing
from the cloud to the network edges that are close to both producers and
consumers~\cite{ananthanarayanan2017real}. This
paradigm shift is generally labelled fog or edge computing,
which involves computing in both cloud and edge environments~\cite{satyanarayanan2017emergence}.
For consistency we will use the term \emph{fog computing} throughout the paper. 

Various fog computing frameworks exist, such as Azure IoT Edge~\cite{IoTEdge},
AWS Greengrass~\cite{Greengrass}, EdgeX~\cite{EdgeX}, and Baidu
OpenEdge~\cite{OpenEdge}. However, their programmability for IoT services is
limited in terms of flexibility due to the below reasons.

\begin{enumerate}[leftmargin=*]
    \item The design and deployment of services is \emph{bound to specific edge
        devices}. This is an edge-oriented approach 
        and requires service developers to statically define which
        service module should be deployed on which type of edge device. The set
        of service modules running on an edge device is fixed after deployment.
        However, for situation-aware IoT applications, different service
        modules need to be triggered dynamically at the network edge according
        to the availability and mobility of IoT devices. This requires \emph{a
        data-oriented approach}.
    \item Existing fog computing frameworks have \emph{poor support for
        data-intensive IoT services.} For example, most existing fog computing
        frameworks use a topic-based pub/sub interface, such as MQTT, for
        communication between different edge service modules and require a
        manual configuration of the data routing path.
        This is problematic when a running task instance at the edge needs to
        be migrated to the cloud or another edge because of
        device mobility, workload fluctuation or when we need to add or
        remove service modules dynamically due to changing business needs.
        Therefore, \emph{IoT services require a dynamic service composition.}
\end{enumerate}

Recently, \emph{serverless computing}~\cite{ServerlessViewBerkeley} is promoted
by major cloud providers to support Function-as-a-Service (FaaS) computing in a
lightweight, dynamic and event-driven manner. At first glance, this makes it
a good fit for the dynamic characteristics of fog computing. However,
serverless computing frameworks only deal with the execution management of
functions, completely separating it from data management. This separation
benefits the simplicity of serverless computing, but has drawbacks for
data-intensive batch and stream processing
applications~\cite{HellersteinCIDR19, atc18serverless}. 
For fog computing, where data locality is paramount, this separation is a deal
breaker and results in poor system efficiency.

In this paper, we design and implement serverless fog computing to support
data-centric IoT services in an edge-cloud environment. We address the
limitations of existing serverless computing and fog computing frameworks,
improving their efficiency and flexibility. Specifically, we propose a fog
function programming model and a context-driven orchestration runtime system to
enable serverless fog computing. Our main technical contributions are listed as
follows:

\begin{figure*}[t]
    \centering
    \includegraphics[width=0.90\linewidth]{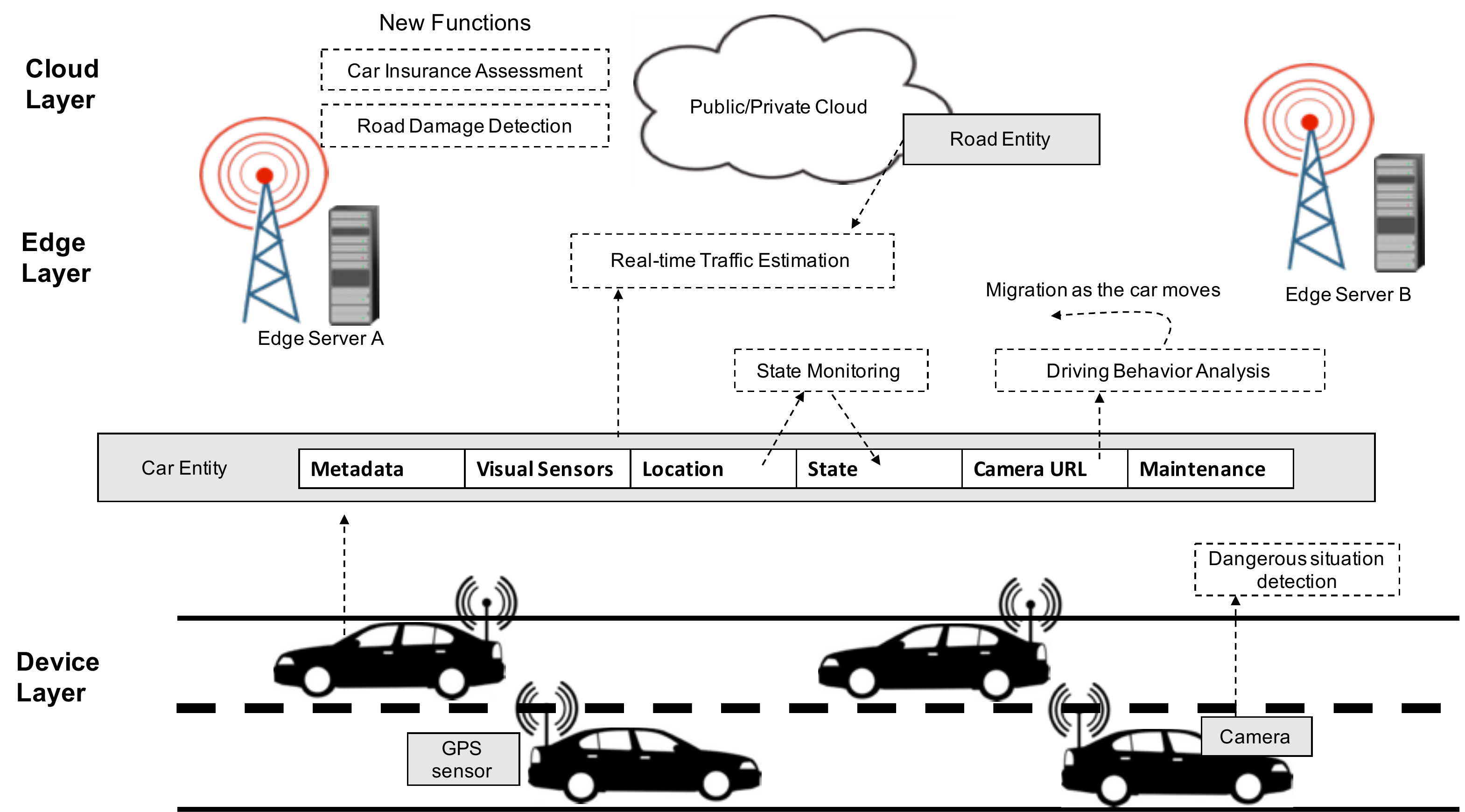}
    \caption{The connected car use case.}
    \label{fig:motivation-use-case}
\end{figure*}

\begin{itemize}[leftmargin=*]

    \item We design \emph{Fog Function} as an enhanced Function-as-a-Service
        programming model, which relaxes the lifetime and resource constraints
        of traditional cloud functions and allows the seamless move from code
        to data or from data to code. Based on the validation and analysis of
        two use cases, we show that Fog Function is more flexible
        and efficient than traditional event-based cloud functions to support
        serverless fog computing for data-intensive IoT services.

    \item We propose a context-driven orchestration mechanism to dynamically and
        automatically trigger, configure, and optimize the deployment of Fog
        Function in cloud and edge environments. Our approach not only
        supports mobility aware task migration and but also achieves better load
        balance across edge nodes as compared to edge-oriented
        approaches applied in current fog computing frameworks.  

    \item We introduce the detailed mechanism of task deployment and migration to
        support context-driven Fog Function and report the performance results in
        terms of latency and scalability.  
 
\end{itemize}

\section{Motivation and Gap Analysis}
\label{sec:motivation}

\subsection{Motivating Use Case}

Several domains, such as smart cities, automotive, and smart manufacturing
benefit from serverless fog computing.
For example, with a growing number of connected cars, there is an emerging
demand to deploy IoT services to enhance driving experience and safety by
leveraging the data produced by connected cars and other data sources.
Figure~\ref{fig:motivation-use-case} depicts such scenario with four IoT
services based on data from four entity types: {\em Road, Car, Camera, and GPS
sensor}:

\begin{enumerate}[leftmargin=*]
    \item [\emph{(S1)}] \emph{State monitoring}: This service takes the
        location updates of a car and then checks whether the car is moving in a
        normal state; in the end it updates the state of the car entity. 

\item [\emph{(S2)}] \emph{Driving behavior analysis}: If the car is in an
    unusual state, this function will be triggered for a detailed inspection of
    the driver's behavior, based on the captured image from a camera in the
    car. 

\item [\emph{(S3)}] \emph{Real-time traffic estimation}: Estimate the real-time
    traffic information aggregated at various levels, e.g., per road, per
    district, per city.   

\item [\emph{(S4)}] \emph{Dangerous situation detection}: Detect any dangerous
    situation on the road and then update the road entity in order to inform
    the other drivers behind on the same road. 
\end{enumerate}

\subsection{Diversity and Dynamics of IoT Data and Workload}

In the car use case the diversity and dynamics of IoT data and workload are
reflected by the following observations. Assume that all data are represented
as entities while the workloads of IoT services are represented by data processing tasks.
Each task takes some entities in, performs some internal data processing, and
then produces some outputs to create new entities or update existing entities.

\noindent
\emph{(O1) Small vs.\@ big entity}: Entities processed by IoT services
differ in size. For example, a car entity can contain lots of information about
the car, such as location, manufacture information, embedded sensors, cameras,
maintenance information, and other metadata. A road entity might be small.

\noindent
\emph{(O2) Static vs.\@ dynamic entity}: Some entities are static or do
not change frequently, while others change frequently. E.g., the location of a
car entity changes constantly. 

\noindent
\emph{(O3) Small vs.\@ big task}: The required computation per task differs.
For example, the state monitoring task is much more light-weight than the
image-based driving behavior analysis.  

\noindent
\emph{(O4) Short vs.\@ long task}: The lifetime of tasks differs
and is often bound to the availability of input data. 
For example, the driving behavior analysis is only triggered if the car entity state is set to
be ``abnormal'' by the state monitoring task.  

\noindent
\emph{(O5) Normal vs.\@ urgent task}: Different tasks might have different
priorities. For example, S4 has higher priority than all the other services and
it needs to have more resource or even exclusive resource usage at computing
edges since the available resource at each edge is usually limited. 

\noindent
\emph{(O6) Existing vs.\@ new task}: New services might be added on the fly
at runtime. For example, road damage detection can be deployed to assess the
road condition and then trigger timely road inspection and maintenance; suspect
detection can be launched by the law enforcement office to search and track
suspicious attackers during an emergency situation. These new tasks need to
reuse raw data published by devices and intermediate results generated by other existing tasks.   

\subsection{Gap Analysis}

To realise these IoT services the following two expectations need to be met:
First, in the design phase, service providers want to have the flexibility to
add, remove, update, and compose services on the fly as their business evolves
over time. Second, during the operating phase, these services should be able to
run seamlessly and efficiently across geo-distributed clouds and edges managed
by infrastructure providers. To meet both expectations, we identify the
following gaps that a fog computing framework needs to address:

\noindent
\emph{(G1) Data discovery and routing: from topic-based to content-based}: Raw
data and intermediate results should be forwarded to different tasks based
on their needs. Existing fog computing frameworks use a topic-based
pub/sub interface, such as MQTT, to configure the data routing paths between
different tasks. However,
representing all entity data with topics is inefficient because some task might
only need part of the whole entity data (O1). Thus, to efficiently discover and
forward any required data to a task, the management and discovery of IoT data
needs be content-based. 

\noindent
\emph{(G2) Function triggering: from per event to per selected entities}: In
existing serverless computing frameworks, functions are invoked per event with
limited execution time and memory size. This is not suitable for data-intensive
IoT services (O3 and O4). First, it is difficult
for service designers to know the execution time and memory size required by a
function during the design phase.
Second, the input of a function can be a single entity update or a stream of updates (O2)
and lifetime is usually associated with the availability of input data.
Triggering a function with the availability of its inputs not only avoids the
effort to explicitly define the triggering event, but also allows service
developers to follow data-centric design principles.  

\noindent
\emph{(G3) Function execution: from data$\rightarrow$code or
code$\rightarrow$data to code$\leftrightarrow$data}: 
Existing serverless computing frameworks separate data management from the
function execution environment, always moving data into
the execution environment for function execution (data$\rightarrow$code pattern).
On the other hand, existing fog computing frameworks such as Azure IoT Edge
move cloud functions to the data located at the edges. This follows a
code$\rightarrow$data pattern. With regards to our observations O2, O3, and O4, the
workload of various IoT data processing tasks is highly diverse and changes
over time, requiring a dynamic and transparent placement of data and code.

\noindent
\emph{(G4) Function composition: from event-oriented or edge-oriented to
data-centric}: Observation O6 indicates a strong need for function composition.
In existing serverless computing frameworks such as
OpenWhisk~\cite{baldini2016cloud}, service developers need to customize a
series of event triggers and rules to link multiple functions together.
Existing fog computing frameworks allow service developers to link functions at
each edge by manually configuring the topic-based data routing path between
them. However, for data-intensive IoT services, a data-centric approach is more
efficient and flexible because it can directly take advantage of the data
dependency between different functions and perform function composition with a
global view of the entire data layer, rather than a subview of each edge.

\section{Fog Function for Serverless Fog Computing}
\label{section:design}

To fill the aforementioned gaps, we propose a new data-centric function
programming model called \emph{Fog Function}. We also introduce the underlying
context-driven service orchestration mechanism of the programming model. 

\subsection{High Level System View}

\begin{figure}[h]
\centerline{\includegraphics[width=0.8\linewidth]{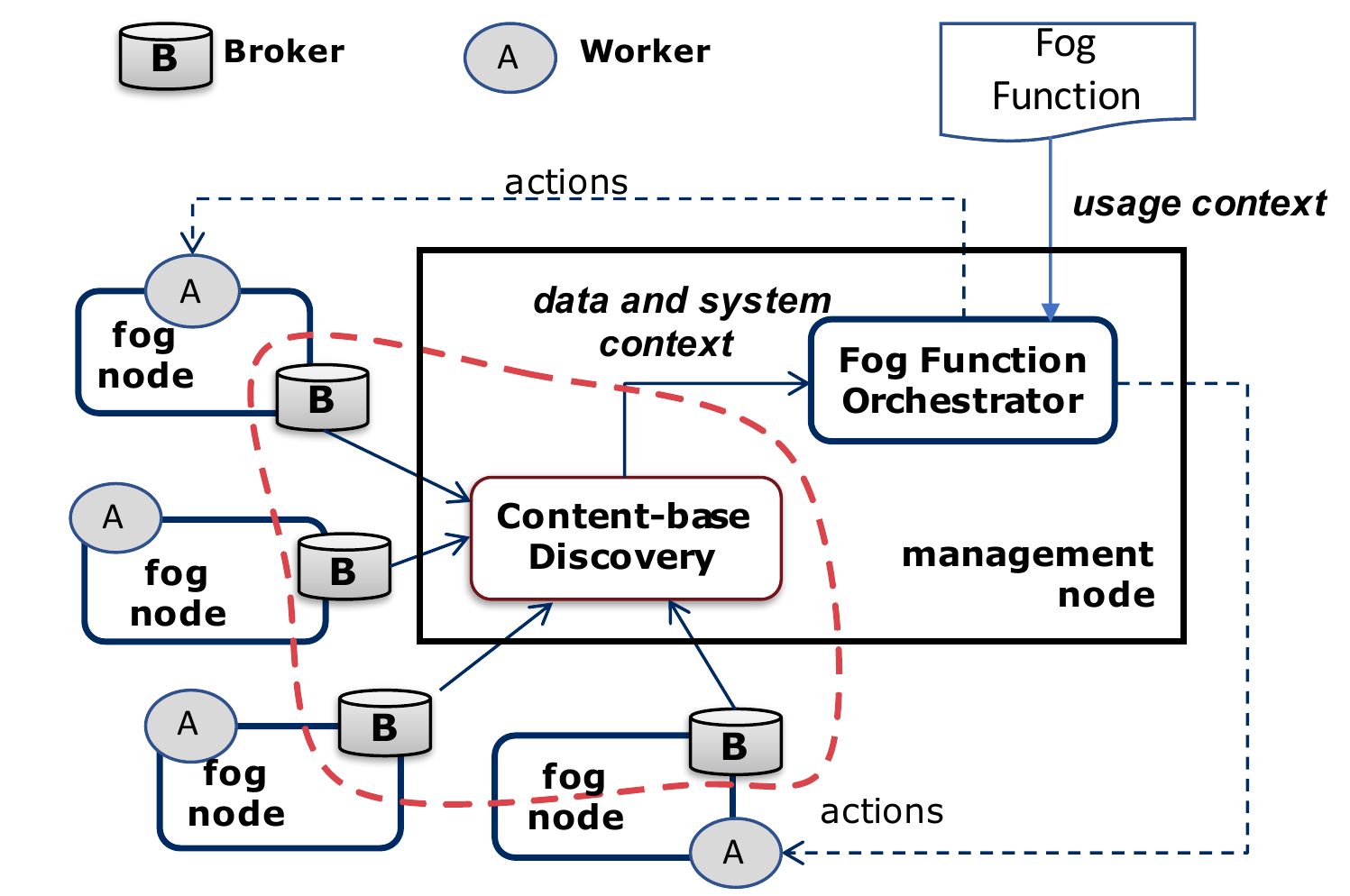}}
\caption{High level system view}
\label{fig:systemview}
\end{figure}

Figure~\ref{fig:systemview} shows the high level view of our system for the
orchestration of fog functions. The system consists of a number of \emph{fog
nodes}, each of which runs a \emph{Broker} and a \emph{Worker}. A
management node runs two centralized components, namely \emph{Discovery} and
\emph{Orchestrator}. Each node is a Virtual Machine (VM) or physical host deployed either in the
cloud or at edges. All fog nodes form a hierarchical overlay based on
their configured GeoHash IDs. All data in the system are represented as
entities saved by a Broker and indexed by the centralized Discovery for
discovery. The data can be raw data published by IoT devices, 
intermediate results generated by some running data processing tasks, 
or available resource data reported by fog nodes. When a fog function is registered,
Orchestrator will subscribe to the input data of the fog function to Discovery.
Once the subscribed data appear or disappear in the system, Orchestrator will
be informed and then take orchestration actions accordingly, which will be
carried out by an assigned worker.   

\begin{figure}[h]
\centerline{\includegraphics[width=0.8\linewidth]{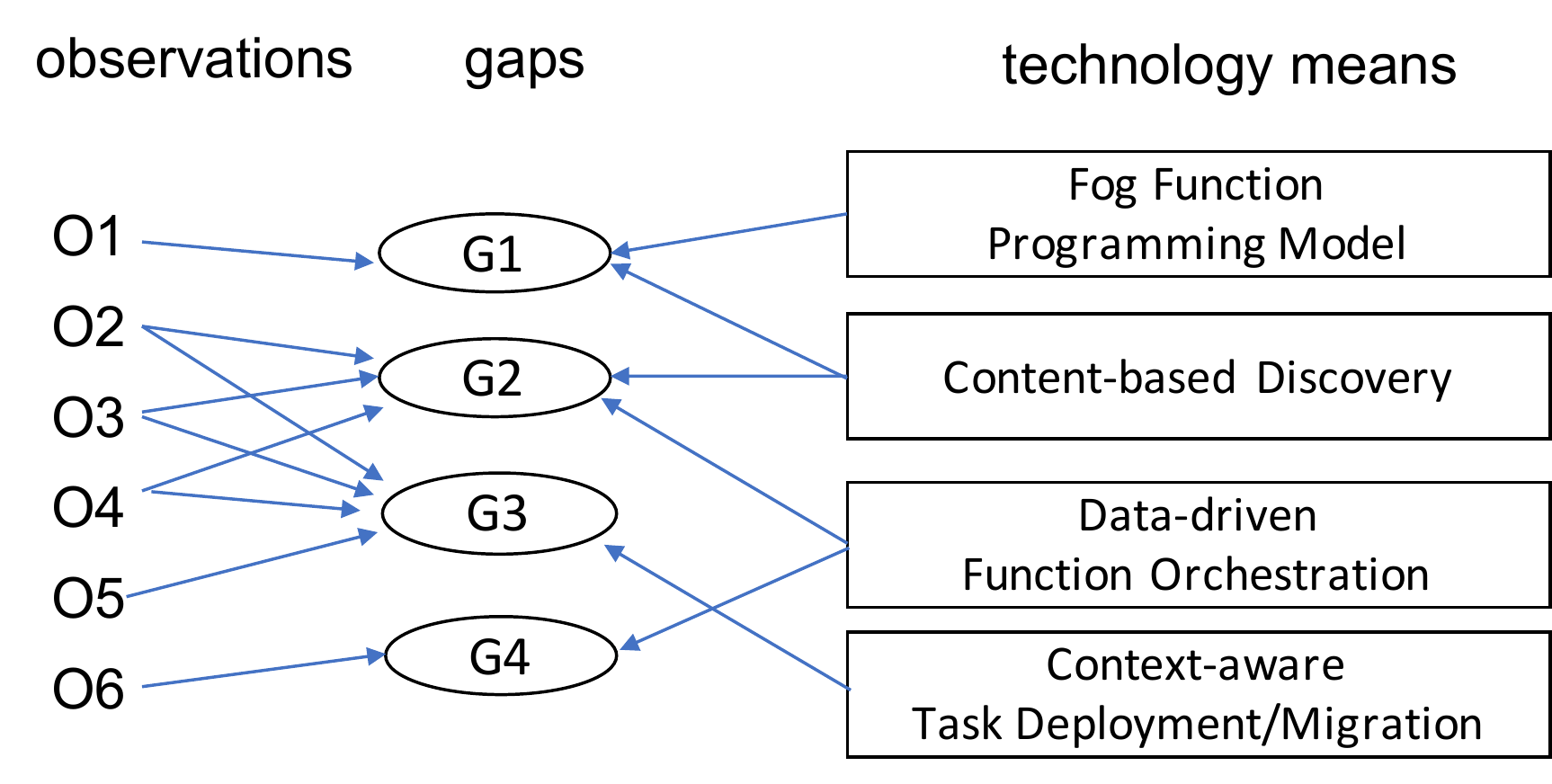}}
\caption{The mapping between gaps and observations/proposed means}
\label{fig:mapping}
\end{figure}

Notice that with our approach the orchestration is also driven by events 
but limited to only two pre-defined events: entity ``appear'' or ``disappear''.
It also provides a declarative interface for service designers to easily annotate
which data should be used to trigger which function with a customizable granularity 
and some other high level orchestration intentions. 
Therefore, service designers can fully focus on the data aspect without explicitly defining the triggering events. 
Figure~\ref{fig:mapping} shows the relationship between the identified gaps and our proposed technology. 
More details are provided in the following sections. 

\subsection{Fog Function Programming Model }

Each fog function is presented as an entity as well, but it is annotated with the following attributes. 

\begin{itemize}

\item \emph{Name}: a unique identity of the function. 

\item \emph{Operator}: the name of a data processing operator. The operator is
    implemented as a dockerized application based on the interface of fog
    function described below.  The specified operator is
    instantiated at runtime by a Worker as a task with its configured inputs and outputs.
    The task is deployed in a dedicated Docker container running on a fog node.

\item \emph{Inputs}: a set of selected inputs required by the operator to do internal data processing. 

\item \emph{Outputs}: the entity type generated by the operator. 

\item \emph{Geoscope}: the geoscope to be applied when selecting input data for this fog function. 

\item \emph{Priority}: the priority of this fog function, which will be taken into account by workers to decide 
how to assign their limited resources to different functions at fog nodes.  

\item \emph{SLO}: the expected \textbf{S}ervice \textbf{L}evel
    \textbf{O}bjective, which is defined as various optimization goals, for
    example, minimizing the latency to produce outputs, maximizing the accuracy
    of generated results, or minimizing the bandwidth usage across fog nodes.
    Different SLO leads to different task deployment plans.  

\end{itemize}

The last four attributes are optional. The inputs are the key for
Orchestrator to decide when to trigger the fog function and how to create its
tasks. Each input is further defined with the following information. 

\begin{itemize}

\item \emph{SelectedType}: the entity type of this selected input. 

\item \emph{AttributeSet}: the required attributes of the selected entity. 

\item \emph{Constraints}: the filters to further select input entities based on some specific attribute values. 

\item \emph{GroupBy}: the granularity to control how many tasks should be instantiated 
	and how the selected input entities should be assigned across its tasks. 
	It can be defined as ``per entityID'', ``per entityType'', or ``per attributeValue''. 

\item \emph{Scoped}: this could be true or false and it is used to decide
    whether geoscope should be applied to select input data when the geoscope is defined with the fog function. 

\end{itemize}

The designed interface for developers to program an operator is shown as below.
The first parameter \emph{entity} is the received data for internal processing.
The other parameters, publish, query, and subscribe, 
are the callback functions for the internal function code to interact with the data management layer
via a nearby broker assigned by Discovery, 
such as, publish generated entity data, query or subscribe additional information.

\begin{verbatim}
function(entity, publish, query, subscribe) 
\end{verbatim}

As compared to the existing FaaS programming models 
like \emph{Cloud Function} in existing cloud-based serverless computing frameworks
or \emph{Edge Function} in the existing edge-centric fog computing frameworks, 
Fog Function has a similar interface for developers
to write function code, but it provides some unique annotations for the
underlying runtime system to efficiently trigger, execute, and compose
functions over cloud and edges in a flexible and transparent manner. The
comparison results are summarized in Table~\ref{tab:comparison} from different perspectives. 

\begin{table*}[t!]
\caption{Comparison with existing function-based programming models}
\begin{center}
\begin{tabular}{|c|c|c|c|}
\hline
  &\multicolumn{3}{|c|}{Function-based programming models} \\
\cline{2-4} 
Differentiator & \textit{Cloud Function}& \textit{Edge Function}& \textbf{\textit{Fog Function}} \\
\hline
Execution environment& centralized cloud & each edge & cloud and edges \\
\hline
Input and output bindings & one event  & per topic & selected entities \\
\hline
Configuration &  none & none & tunable parameters \\
\hline
Task granularity &  none & none & definable \\
\hline
Trigger & per event  & per edge &  availability of selected entities\\
\hline
Execution pattern & data $\rightarrow$ code & code $\rightarrow$ data & code $\leftrightarrow$ data \\
\hline
Migration & yes & no & yes \\
\hline
Priority & none  & none &  yes \\
\hline
Service level objective &  none & none &  definable \\
\hline
\end{tabular}
\label{tab:comparison}
\end{center}
\end{table*}

\subsection{Content-based Discovery}

\begin{figure}[h]
\centerline{\includegraphics[width=\linewidth]{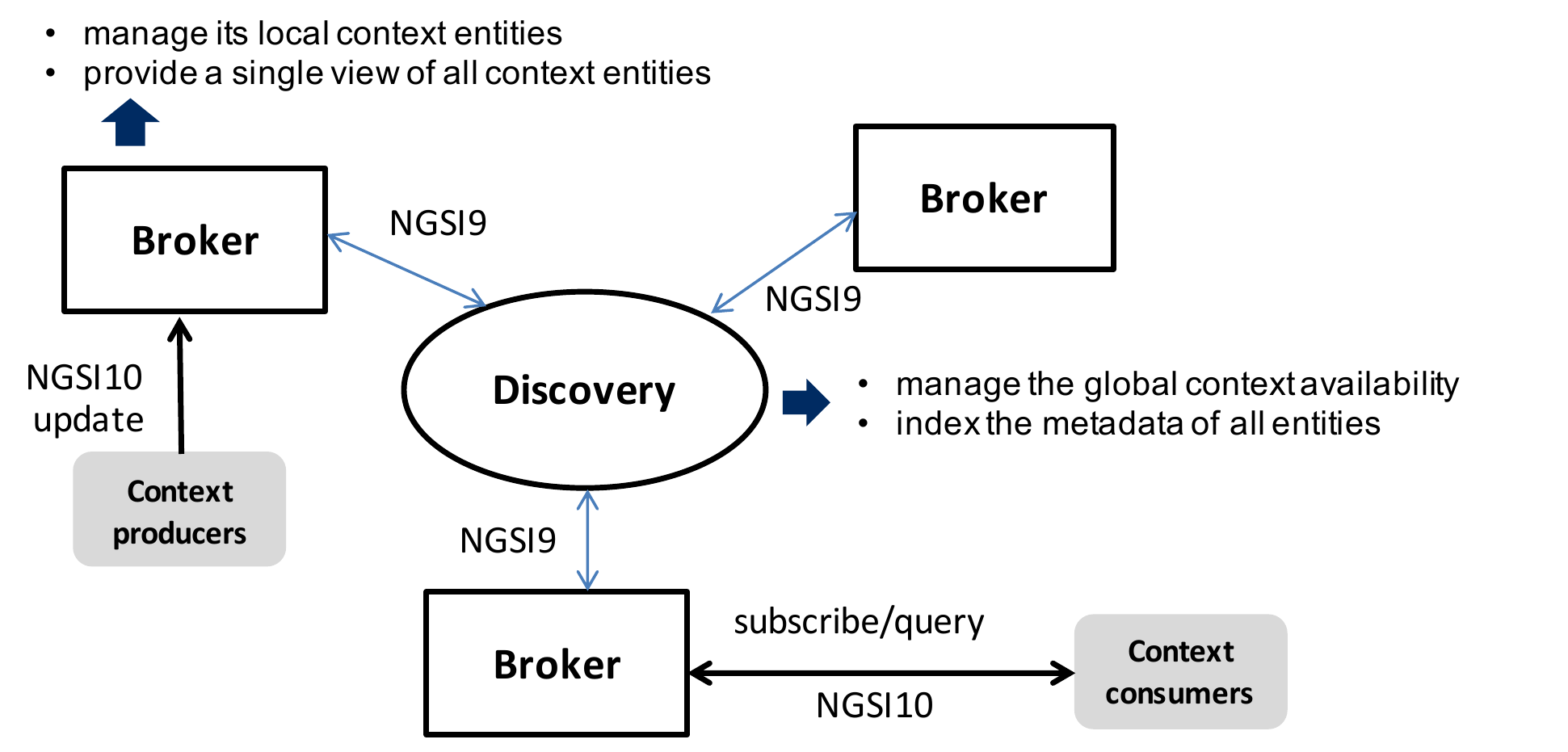}}
\caption{Context discovery with two-layer design}
\label{fig:discovery}
\end{figure}

The context management layer consists of a network of Broker(s) and a centralized Discovery. 
It is designed to provide a global view for all system components and running
tasks to query, subscribe, and update context entities via the unified and standardized data
model and communication protocol, namely NGSI~\cite{NGSIMartin}. 
It plays a very important role to support the orchestration of Fog Function. 
As illustrated by Figure~\ref{fig:discovery}, in our design a large number of distributed Brokers work in parallel 
under the coordination of the centralized Discovery. 
The Discovery component can be used to discover both devices, intermediate data, and available resources at fog nodes as well. 

As compared to the topic-based data management in existing systems like MQTT-based Mosquitto or Apache Kafka, 
our two-layer context management design has the following features: 
1) separating context entity data and their availability; 
2) providing separated and standardized interfaces to manage both context data
(via NGSI10~\cite{NGSIMartin}) and context availability (via NGSI9~\cite{NGSIMartin}); 
3) supporting not only ID-based and topic-based query and subscription but also geoscope-based or attribute-based query and subscription. 

\subsection{Data-driven Function Orchestration}

\begin{figure}[bh]
\centerline{\includegraphics[width=0.8\linewidth]{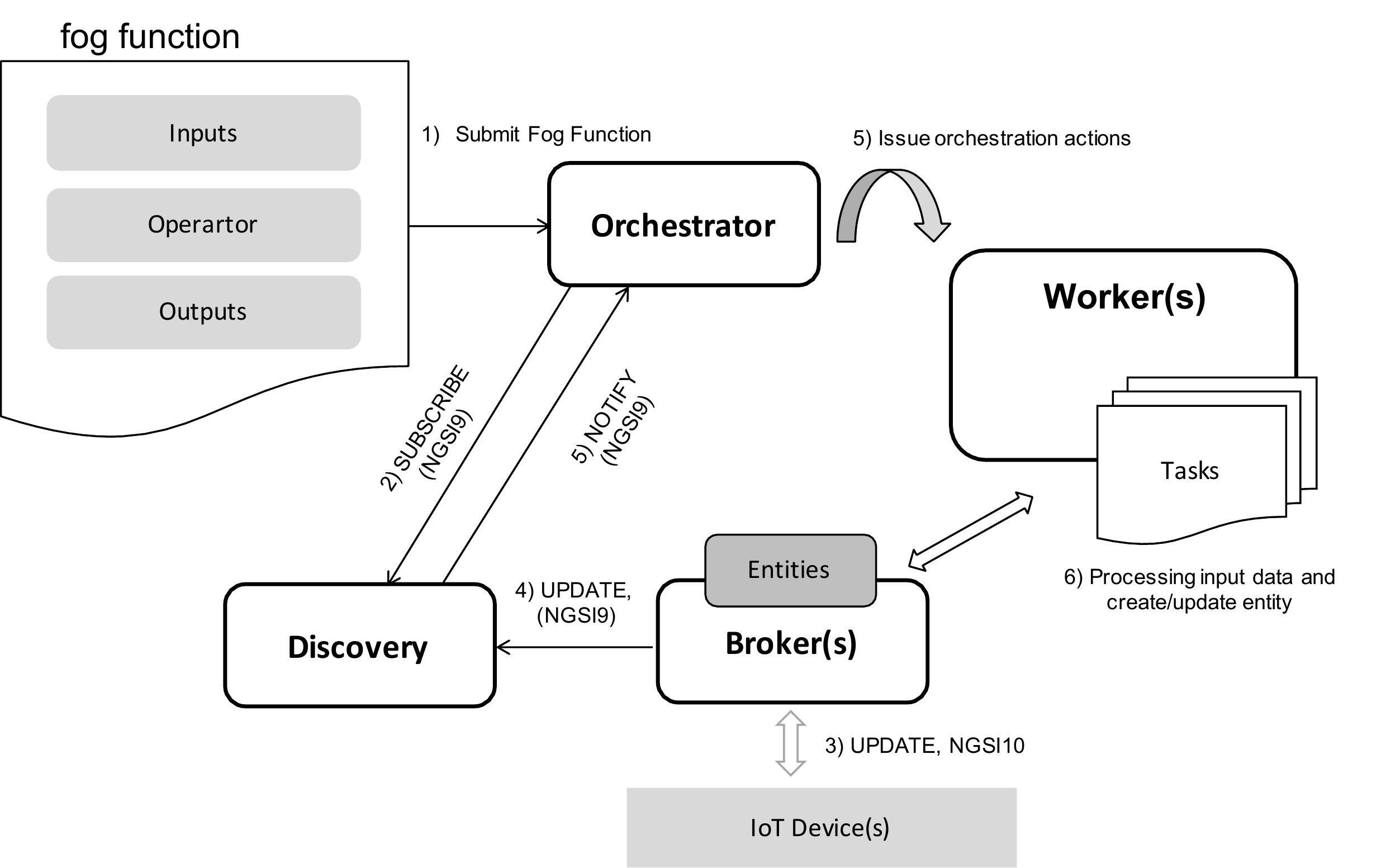}}
\caption{Data-driven orchestration}
\label{fig:orchestration}
\end{figure}

Figurer ~\ref{fig:orchestration} shows the major procedure for Orchestrator to orchestrate fog functions 
based on the update notification of context availability of their input data, provided by Content-based Discovery. 
More specifically, the following four basic orchestration actions are designed to dynamically orchestrate tasks for each registered fog function.  

\begin{itemize}
\item \emph{ADD\_TASK}: To launch a new task with the given
    configuration that includes the initial setting of its input streams. When
    launching a new task, the Worker first fetches the Docker image for this
    task and then launches and configures this task within a dedicated Docker
    container. After that, the Worker subscribes the input entity to the
    context management system on behalf of the running task so that the input
    streams can be received by the running task; 
in the end, the newly created task is reported back to the orchestrator. 

\item \emph{REMOVE\_TASK}: To terminate an existing running task with the given task ID. 
When terminating an existing task, the Worker not only stops and removes its corresponding Docker container, 
but also unsubscribes its input streams so that the context management system does not end up with lots of unavailable subscribers. 

\item \emph{ADD\_INPUT}: To subscribe to a new input stream on behalf of a running task so that the new input stream can flow into the running task. 

\item \emph{REMOVE\_INPUT}: To unsubscribe from some existing input stream on behalf of a running task so that the task
stops receiving entity updates from this input stream. 
\end{itemize}

\subsection{Context-aware Task Deployment/Migration}

Overall the orchestration of Fog Function leverages the following three types of context information.  

\emph{Data context}: the structure and registered metadata of available data,
including both raw sensor data and intermediate data. 
Based on the standardized and unified data model and communication interface, namely NGSI,
our system is able to see the content of all data generated by sensors and data
processing tasks in the system, such as data type, attributes, registered
metadata, relations, and geo-locations. 

\emph{System context}: available resources at each fog node. The resources in a
cloud-edge environment are geo-distributed and they are dynamically changing
over time. As compared to cloud computing, resources in such a cloud-edge
environment are more heterogeneous and dynamic.

\emph{Usage context}: high level usage intentions defined by service designers
to indicate what their fog functions should be used in the system, such as
which type of results is expected under which type of QoS within which
geo-scope.

Task migration is the combination of removing an existing task on Worker A and adding a new task on another Worker B. Currently, we only support the migration of stateless tasks, meaning that the tasks do not hold any persistent internal state and terminating or restarting them does not lead to any faulty state or result. More specifically, our design can allow seamless task migration across cloud and edges in the following three cases. 

\begin{itemize}
\item \emph{Cloud $\rightarrow$ Edge}: for example, if a new edge node joins the system and it is close to one edge device, a task running in the cloud can be migrated from the central cloud to this new edge node in order to save bandwidth. 

\item \emph{Edge $\rightarrow$ Cloud}: for example, when an edge node becomes overloaded since it has to handle the workload from lots of devices in the same region, it can start to migrate some existing tasks to the cloud in order to keep enough resource for the other urgent tasks. 

\item \emph{Edge $\rightarrow$ Edge}: for example, when a mobile device such as a connected car moves from one region (R1) to another region (R2), it might find another nearby edge node that can provide lower latency. In this case, it is better to migrate the task to the edge node in R2. In this case, tasks are migrated in order to adapt to the movement of the mobile devices. 
\end{itemize}

\section{Implementation and Use Case Validation}
\label{sec:validation}

Fog Function has been applied into the open source fog computing framework
FogFlow as a new programming model to enhance its programmability. Originally,
FogFlow can orchestrate dynamic data flows over cloud and edges using a service
topology ~\cite{FogFlow}. The service topology statically defines the logical data processing
flow of an IoT service and is triggered on demand by the requests from the
consumer side, but it does not support the composition of multiple service
topologies and it is not flexible to handle use case requirements which may
change over time. Unlike service topology, Fog Function is simple and flexible
and it is triggered when its input data becomes available. FogFlow can
automatically chain different functions and allows more than one Fog
Function to handle new data items. In the end, the entire execution graph
can be automatically triggered, composed, and managed as data arrives. From the
design perspective, Fog Function is more flexible than the service topology,
because the overall processing logic of an IoT service can be easily changed
over time by adding or removing functions when the service processing logic
needs to be modified for new use case requirements. 

Using Fog Function, we are able to easily realize a {\em smart parking} use
case, which was difficult to achieve with the service topology programming
model. This use case is implemented together with our European project partner,
University of Murcia, based on the real scenario of Murcia City. In Murcia,
there are two types of parking sites, regulated parking zones that are operated
by the city government and can provide historical information of how parking
slots are used per day, and private parking sites that are operated by private
companies and can provide real-time availability of parking spots. By utilizing
these two types of data sources and other public transportation information,
our smart parking service can provide real-time and personalized parking
recommendations for drivers.

\begin{figure}[h]
\centerline{\includegraphics[width=0.6\linewidth]{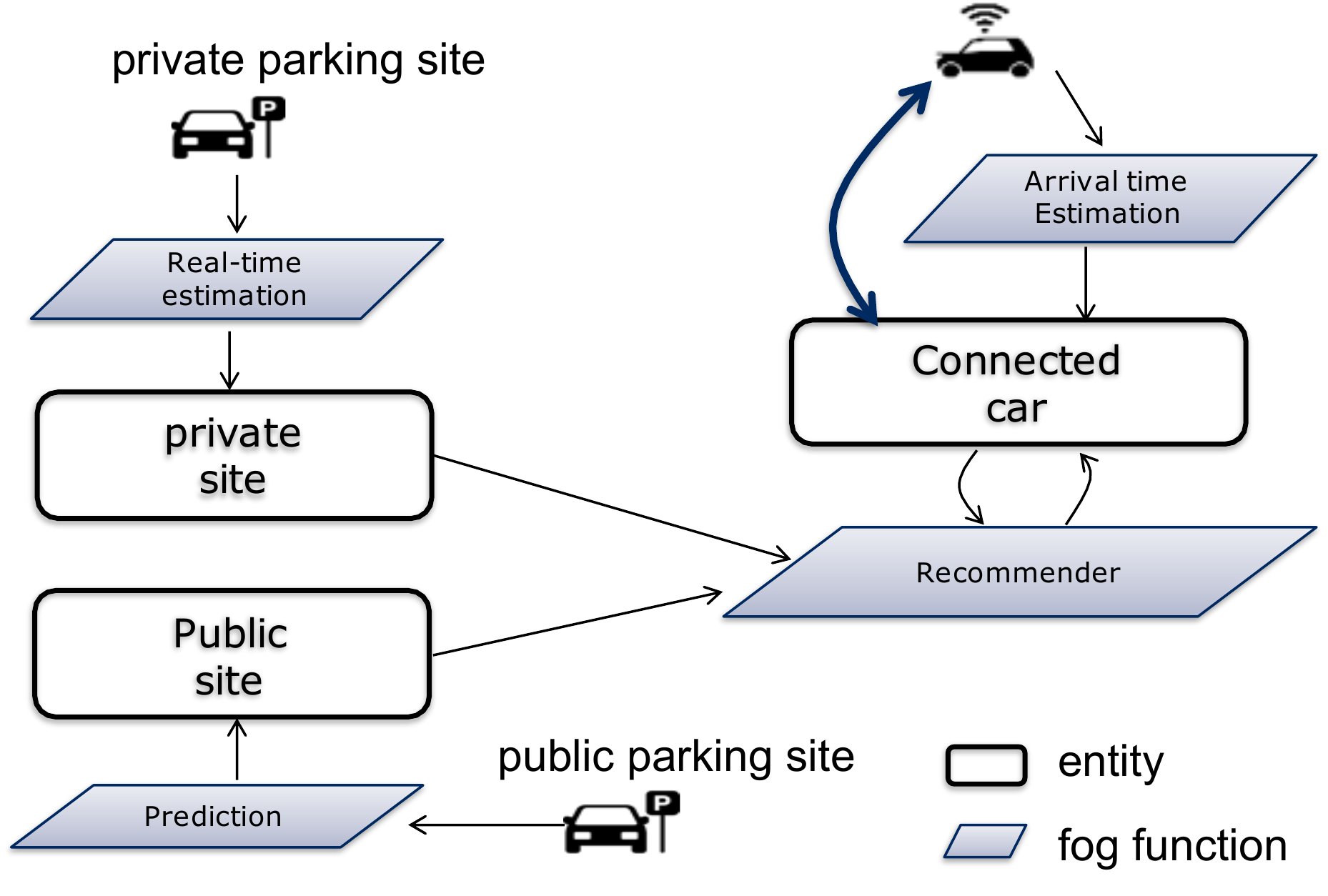}}
\caption{Design of fog functions for a smart parking use case}
\label{fig:smartparking}
\end{figure}

As illustrated in Fig.~\ref{fig:smartparking}, we just need to design and
implement dedicated fog functions for each physical object involved in the use
case. For example, one fog function for each public site to predict how many
parking spots are available per 10 minutes based on their historical
information; two fog functions for each connected car, one to estimate its
arrival time according to the traffic situation on the way and the other to
calculate at which park site the driver can get a parking spot on arrival. The
deployment of those fog function instances are on the edge node close to their
input data sources so that FogFlow can reduce more than 50\% bandwidth
consumption and also provide real-time parking recommendation for each driver.

\section{Performance Evaluation}
\label{sec:evaluation}

Our performance evaluation has been conducted by using a set of virtual
machines from Google Cloud. The whole system setting includes 3 parts: 
1) a client machine that can simulate a set of IoT devices; 
2) a set of fog nodes, of which each is a standard VM with 2 vCPU and 7.5 GB memory (``n1-standard-2''); 
3) the cloud part that includes just one more powerful VM with 8 vCPUs and 7.2 GB memory (``n1-highcpu-8''), 
running the two centralized components, namely Orchestrator and Discovery. 
To trigger the designed data processing tasks defined by service topology or Fog Function, we simulate a set of IoT devices for each test case.  For each test case, we carry out 10 runs of tests. 
In each test, we start a number of simulated devices and keep them running for 10 minutes.

\begin{figure*}[t]
    \centering
    \begin{subfigure}[t]{0.3\linewidth} 
        \centering
        \includegraphics[width=\textwidth]{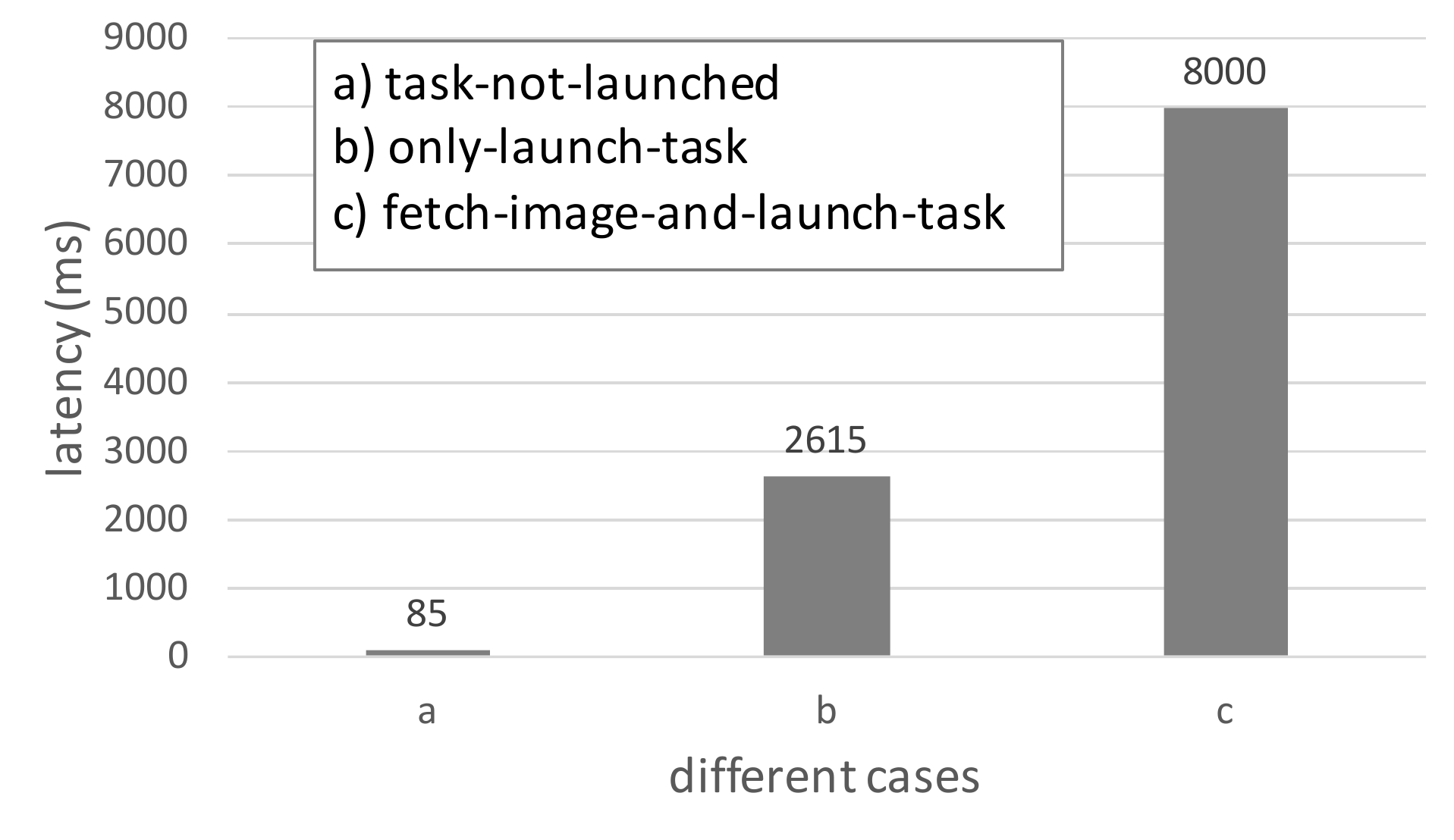}
        \caption{Startup Latency}
        \label{fig:startuplatency}
    \end{subfigure}
    \begin{subfigure}[t]{0.3\linewidth}
        \centering
        \includegraphics[width=\textwidth]{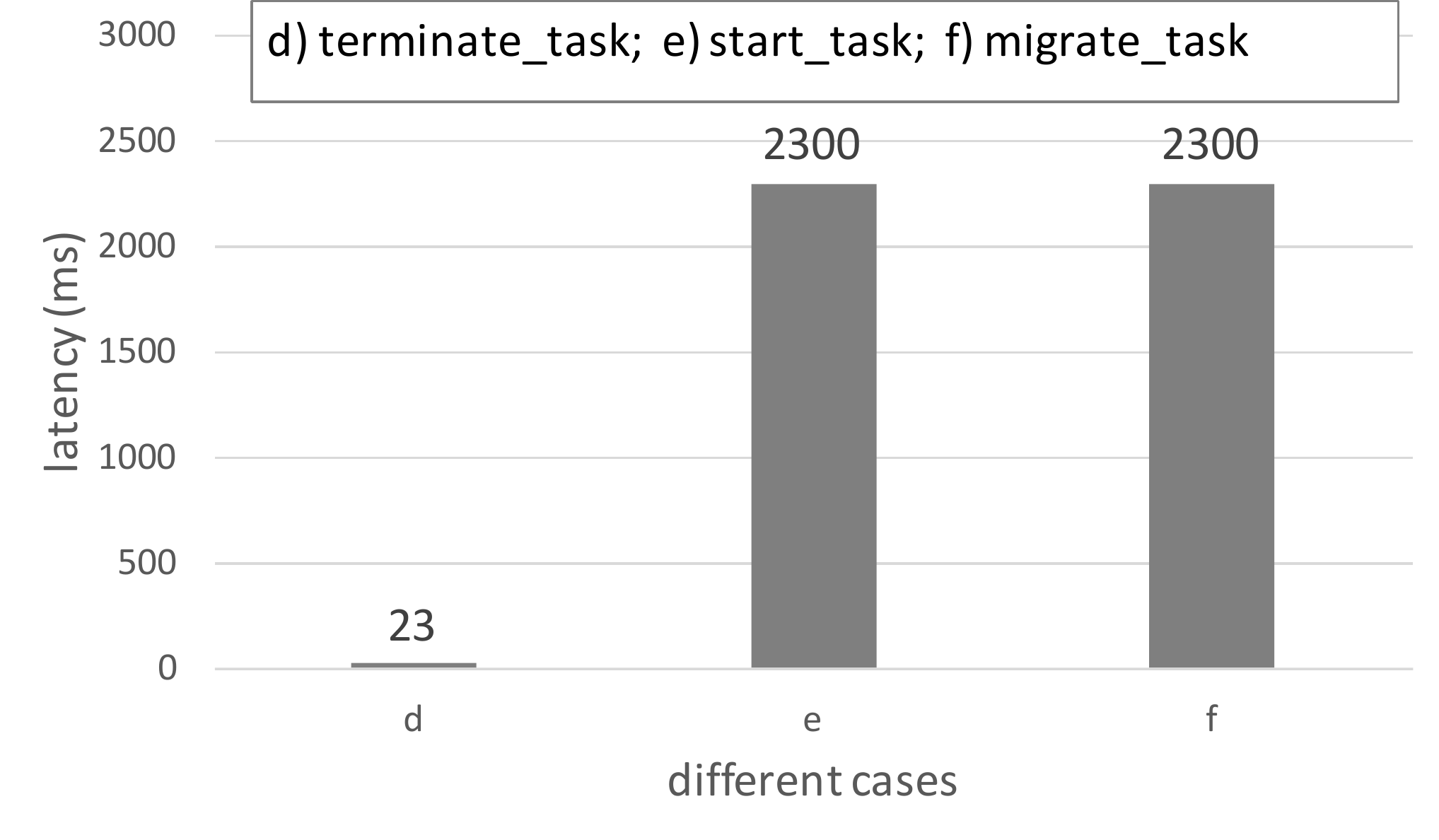}
        \caption{Migration Latency}
        \label{fig:migrationlatency}
    \end{subfigure}
    \begin{subfigure}[t]{0.3\linewidth}
        \centering
        \includegraphics[width=\textwidth]{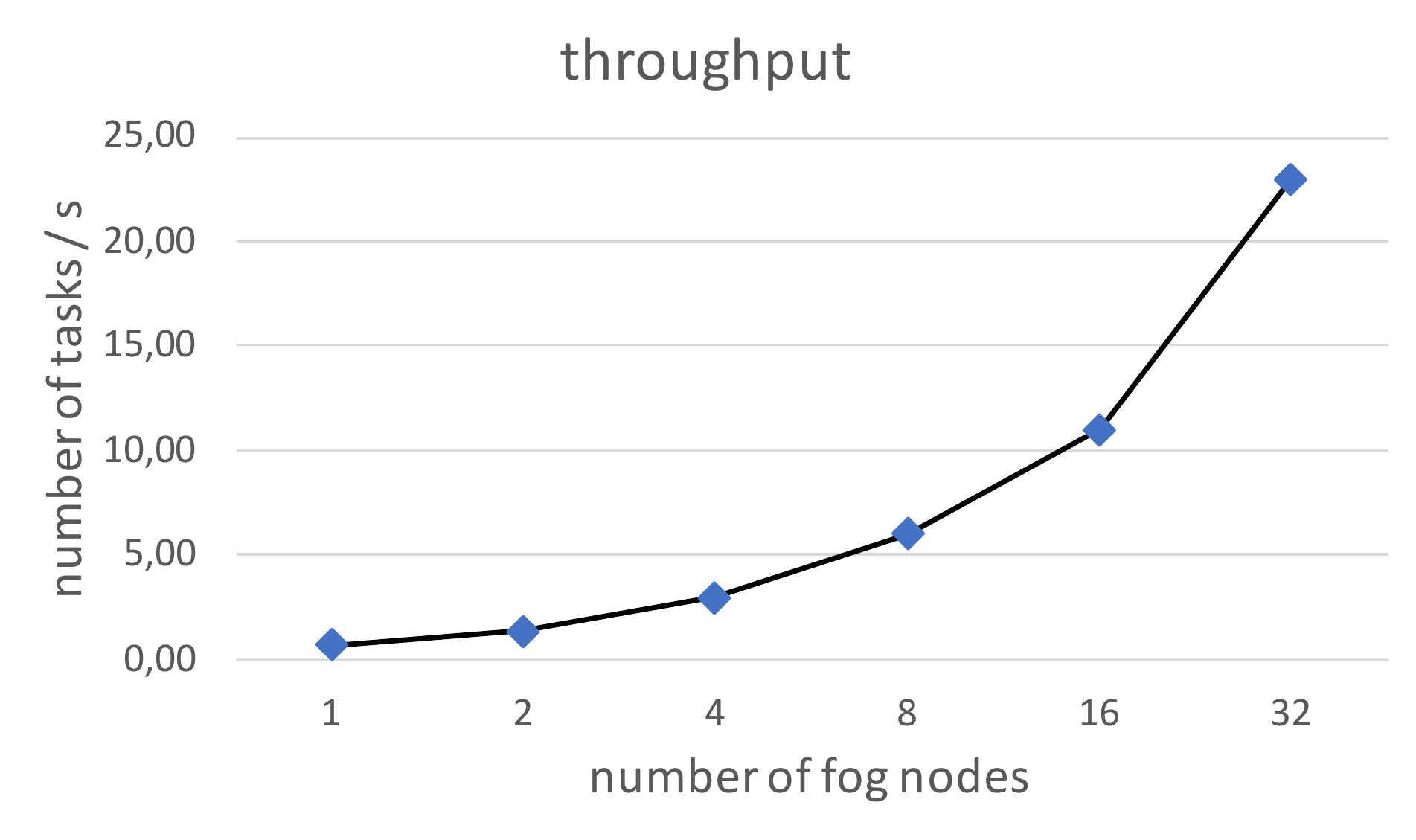}
        \caption{Throughput of Orchestrator}
        \label{fig:throughput}
    \end{subfigure}    
    \caption{Initial measurement results of the FogFlow system}
    \label{fig:measurement}
\end{figure*}

\subsection{Latency}

Figure~\ref{fig:startuplatency} shows the results of startup latency measured in the following three scenarios: 
1) without launching the actual task (named as ``task-not-launched''); 
2) the docker image of the dummy task is not fetched in advanced (named as ``fetch-image-and-launch-task''); 
3) the docker image of the dummy task is already fetched (named as ``only-launch-task''). 
From the measurement result, we can see that the big part of the startup latency 
happens with fetching the docker image from public docker registry. 
Launching a docker container also requires about 2 seconds. 
These two parts of latency are also related to the size of the required docker image. 
The bigger the docker image is, the longer it will take. 
However, the actual time taken by Orchestrator to make its orchestration decisions is very short, less than 100 ms. 
This result indicates that, by leveraging multiple Workers to launch tasks in parallel, 
we can scale up the system easily, even just using a centralized Orchestrator.  

We also measure the latency of migrating one task from one Worker to another Worker. 
By design, migrating a task is done by two orchestration actions: 
terminating the existing task and starting a new task. 
For stateless tasks, these two actions can be carried out in parallel. 
Figure~\ref{fig:migrationlatency} shows the latency results of terminating/starting/migrating a task. 
We can see that the latency of migrating a task is about 2 $\sim$ 2.5 second, 
nearly equal to the latency of starting a task. 
This is because the two orchestration actions (starting a task and terminating a task) are performed in parallel and starting a task is much slower than terminating a task.

\subsection{Scalability}

We also measured how the FogFlow system can be scaled up by adding more fog nodes in terms of throughput. 
The throughput is defined as the average number of launched tasks per second. 
Figure~\ref{fig:throughput} shows the result. 
In the test, we assume that the docker image is already fetched in advanced. 
From the result, we can see that the throughput increases linearly with the number of fog nodes. 
With 10 fog nodes, FogFlow can handle $\sim$8 tasks per second. 
As seen in Figure~\ref{fig:throughput}, the main bottleneck is actually the underlying docker engine, 
which can only launch 1$\sim$3 docker containers per second. 
A single Orchestrator can handle the orchestration decisions for at least 32 FogFlow Workers. 
In the future, we will utilize the unikernel-based visualization technology, such as Unikraft~\cite{manco2017my}, 
to avoid this bottleneck. 
In addition, to further scale up the system, we need to decentralize the Orchestrator and Discovery as well. 

\subsection{Efficiency}

We evaluate system efficiency for the Fog Function based approach in terms of cross-node traffic and service latency
and compare it with the other two approaches: Cloud Function and Edge Function. 
Cloud Function represents the existing severless computing approach for the cloud environment.
With this approach function is triggered by event based on topic and the data management environment is separated from the execution environment. 
Edge Function represents the existing fog computing frameworks that program services per edge based on topic. 
We change the system setting of FogFlow and the specification of Fog Function to simulate both Cloud Function and Edge Function. For topic-based pub/sub, we use type-based and subscribe to the entire entity; 
for the separation of data management and function execution, we deploy brokers with Discovery and workers with Orchestrator within two groups of VMs. 
We simulate 1000 connected car entities with two types of entity lengths (126 bytes for the small size case and 1682 bytes for the big size case) to trigger a simple speed estimation function. 
Each car reports its current location every second and each run of test is 10 minutes. 
The cross-node traffic is the total amount data transferred across VMs and
the service latency is the delay from when the raw data are published to when the result is produced. 
The comparison results are shown in Table~\ref{tab:efficiency}. 
As compared to Cloud Function, Fog Function can reduce more than 95\% cross-node traffic and about 80\% service latency by leveraging data locality. 
As compared to Edge Function, Fog Function can reduce 30\% service latency for dealing with big entities
thanks to the task migration mechanism, but it slightly introduces 5\% additional cross-node traffic. 
For time-sensitive services, this improvement is important. Also when considering more dynamic services, 
the benefit of Fog Function could be even more clear because Edge Function will suffer from workload imbalance.

\begin{table}[t!]
\caption{Comparison with existing function-based programming models}
  \begin{tabular}{lS|S|S|S}
    \toprule
    Approaches &
      \multicolumn{2}{c}{cross-node traffic (MB)} &
      \multicolumn{2}{c}{avg. service latency (ms)}  \\
      & {small} & {big} & {small} & {big} \\
    \hline      
    Cloud Function & 86.6 & 987.3 & 262 & 610   \\
    \hline
    Edge Function & 3.5 &  10.8 & 68 & 150    \\
    \hline
    Fog Function & 3.8 &  11.4& 59  &  102    \\
    \bottomrule
  \end{tabular}
\label{tab:efficiency}
\end{table}

\section{Related Work}
\label{sec:relatedwork}

\subsection{Fog/Edge Computing}
There are already various fog computing or edge computing frameworks and approaches. 
Their programming models and orchestration mechanisms are defined at different levels, 
from the lower layer infrastructure level to the upper layer application level. 
For instance, Cloudlet~\cite{satyanarayanan2009case} proposes an edge computing approach 
to offload computation from mobile devices to the network edge using virtual machine (VM) based cloudlets. 
KubeEdge~\cite{KubeEdge} is an open source system extending native containerized application orchestration and device management to hosts at the edge. 
Telcofog~\cite{vilalta2017telcofog} defines a unified fog and cloud computing infrastructure for 5G networks over distributed clouds based on both VMs and containers. 
These frameworks are suitable for generic standalone applications that can be hosted either in the cloud or at edges, but they lack a programming model to program the logic of data-intensive applications. 
To overcome this issue, many dataflow-based approaches are proposed.  For example, the initial version of FogFlow~\cite{FogFlow} can program IoT services over cloud and edges based on service topology.  AWS Step Function~\cite{AWSStepFunction} is able to build distributed applications using visual workflows. 
These frameworks are flexible to support the function composition within a single application, 
but they require users to manually trigger the defined application and also the cross-applications function composition is not possible. 
To address these limitations, the function-based programming model is introduced by many fog computing frameworks, 
such as Amazon Greengrass~\cite{Greengrass}, Azure IoT Edge~\cite{IoTEdge}, and Baidu OpenEdge~\cite{OpenEdge}. 
They all trigger functions per edge based on a topic-based pub/sub system. 
Opposed to that, our design can automatically trigger functions for user-definable data granularity based on the availability of their input data.  

\subsection{Serverless Computing}
Serverless computing is emerging as a new paradigm for the deployment of cloud applications. 
Many of the major cloud vendors, including Amazon Lambda, Google Cloud Functions, Microsoft Azure Functions, IBM Cloud Functions, have released their serverless computing platforms. 
In addition, there are also many open source serverless computing frameworks, such as OpenWhisk, Kubeless, Fission, and OpenFaaS. They are all designed for the cloud environment in which computation resource and storage resource are unlimited and centralized. Some studies address the cold start issue of function provisioning~\cite{atc18sand,atc18sock}. On the other hand, as serverless computing is applied into data-intensive applications~\cite{PyWren, ishakian2018serving},  database~\cite{schleier2019serverless}, video analysis~\cite{ao2018sprocket}, the data management and sharing between serverless tasks turns to be a bottleneck. Pocket~\cite{klimovic2018pocket} proposes a fast storage system for ephemeral data sharing between serverless tasks, yet it is still not able to benefit from data locality since the management of data and tasks is separated. Our design can manage data and serverless tasks jointly to achieve better efficiency in more geo-distributed and heterogeneous environments.

\section{Conclusion and Outlook}
\label{sec:conclusion}

In this paper we take a first step into applying the serverless concept into
fog computing for data-intensive IoT services. The goal is to keep the
simplicity and flexibility of the Function-as-a-Service programming model for
fog computing via an extended function programming model called Fog Function,
meanwhile improving the efficiency of existing frameworks with two approaches:
content-based discovery and context-driven orchestration. 
The current approach is scalable with hundreds of fog nodes, but it is
necessary to decentralize the discovery and orchestration for a much larger
scale. Also, in the future the context-driven orchestration mechanism can be
improved to provide predictable service level objectives with some machine
learning based approaches as proposed in~\cite{argerich2019reinforcement}.

\begin{table}[b]
\begin{tabular}{m{0.28\linewidth}m{0.62\linewidth}}
\includegraphics[width=\linewidth]{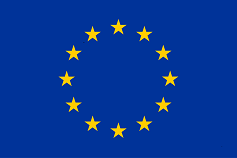} & 
\small{The research leading to these results has received funding from the European Community's Horizon 2020 research and innovation programme under grant agreement n\textsuperscript{o} 779747.}
\end{tabular}
\vspace{-3ex}
\end{table}

\bibliographystyle{IEEEtran}
\bibliography{ref}

\end{document}